\documentclass[12pt,a4paper]{article}

\usepackage{amsmath,amsthm,amssymb}
\usepackage{graphics,graphicx}
\usepackage{epsfig}
\usepackage{multicol}
\usepackage{color}
\makeatletter
\@addtoreset{equation}{section}
\makeatother

\begin{document}

\begin{flushright}
QMUL-PH-06-02
\end{flushright}

\begin{center}
\Large\textbf{Electrified Fuzzy Spheres and Funnels in Curved Backgrounds.}\\
\vspace{2cm}
\normalsize\textbf{Steven Thomas}\footnote{s.thomas@qmul.ac.uk} \normalsize\textbf{and John Ward}\footnote{j.ward@qmul.ac.uk}\\
\vspace{1cm}
\emph{Department of Physics\\
Queen Mary, University of London\\
Mile End Road, London\\
E1 4NS, U. K}
\end{center}
\vspace{1cm}
\begin{abstract}
We use the non-Abelian DBI action to study the dynamics of $N$ coincident $Dp$-branes in an arbitrary curved background,
with the presence of a homogenous world-volume electric field. The solutions are natural extensions of those 
without electric fields, and imply that the spheres will collapse toward zero size. 
We then go on to consider the $D1-D3$ intersection in a curved background and find various dualities and automorphisms of the 
general equations of motion. It is possible to map the dynamical equation of motion to the static one via Wick rotation, however the
additional spatial dependence of the metric prevents this mapping from being invertible. Instead we find that a double Wick rotation leaves the static
equation invariant. This is very different from the behaviour in Minkowski space.
We go on to construct the most general static fuzzy funnel solutions for an arbitrary metric either by solving the
static equations of motion, or by finding configurations which minimise the energy. 
As a consistency check we construct the
Abelian $D3$-brane world-volume theory in the same generic background and find solutions consistent with energy 
minimisation. In the $NS$5-brane background we find time dependent solutions to the equations of motion,
representing a time dependent fuzzy funnel.
These solutions match those obtained from the $D$-string picture to leading order suggesting that the
action in the large $N$ limit does not need corrections. We conclude by generalising our solutions to higher dimensional
fuzzy funnels.
\end{abstract}

\newpage
\section{Introduction.}
The issue of time dependence in string theory has been discussed in a number of recent works, from the boundary
CFT approach and the effective $D$-brane action \cite{time_dependence, costis2}. The most recent efforts have been related to the dynamics of branes
in curved backgrounds with classical supergravity solutions, and shown that there is a similarity between the
brane motion and condensation of open string tachyons. The hope is that understanding of one of these pictures will
lead to better understanding of the other. We would also hope to learn more about the nature of branes with regard to cosmology, which
has been recently dealt with in \cite{cosmological, gibbons, craps}
The obvious objection to this is the fact that branes in the literature are assumed to be rigid
hyperplanes in type II string theory. Whilst this is acceptable from the viewpoint of perturbative string theory, spatial fluctuations of these moving 
branes should be taken into account.

A further thing to note is that almost all of these works have dealt with solitary branes moving in some background,
where we are assuming that there has been some Higgs mechanism employed to separate the branes. Thus a more general
approach would be to consider the dynamics of coincident branes \cite{taylor}, where the massless modes of the theory help define
a $U(N)$ gauge symmetry via the use of the non-Abelian DBI effective action \cite{nonabelianDBI}. 
This is more useful from the viewpoint of cosmology and standard model building, as well as trying to 
understand aspects of black holes physics \cite{gravitationalmyers, diego}.

The key to the non-Abelian DBI relies in the fact that we must employ non-commutative geometry, replacing the
scalar fields (which in the Abelian theory are singlets) by $N \times N$ adjoint valued matrices. At high energies
near the Planck scale we may find that our intuitive ideas about smooth geometries need to be revised, therefore
it is worthwhile to fully understand the nature of non-commutative physics.
Much of the work with the effective non-Abelian DBI has dealt with brane polarisation \cite{nonabelianDBI} and intersections \cite{constable, cook}, however
most of the literature here has also only been concerned with flat backgrounds - and although it is expected
that many of the dualities and brane configurations will hold in curved space, this has not yet been conclusively
proven. Furthermore the flux compactification scenario developed in \cite{KKLT} emphasises the need for more understanding
of systems in warped backgrounds.
Recently there has been suggestion that the non-Abelian action needs to be corrected when dealing
with curved backgrounds \cite{modifiedDBI}. However it was shown in \cite{hyakutake, huang, thomas} that these corrections do not need to be included
when taking the large $N$ limit. This was demonstrated using the gravitational Myers effect \cite{gravitationalmyers}, where the equations of motion
arising from the Abelian and non-Abelian theories are identical to leading order in $1/N$. We should bear
in mind that this may have been a special case, so investigating other non-commutative brane configurations in curved
space should further enhance our knowledge of the symmetrised trace. It may well turn out to be the case that these corrections turn out to
be important when taking the finite $N$ limit. Recently \cite{finiten} have proposed a complete expansion of the symmetrised trace,
which potentially opens up new avenues of investigation.

In this paper we wish to examine some of these issues by extending the research begun in \cite{constable, asano, thomas} to
consider the dynamics of coincident $Dp$-branes in an arbitrary curved background. We will consider the case where
we turn on a homogenous electric field on each of the world-volumes to see how this affects the dynamics, and the
leading order corrections from the application of the symmetrised trace. The main thrust of this work, however, will
concern the brane intersection problem of $D1$-branes with a solitary $D3$-brane in an arbitrary curved background.
It is well known that coincident $D$-strings in flat space 'expand' along their world-volume direction to create
an object known as a fuzzy funnel \cite{constable, cook, koch, berman}. The radius diverges at some point and the configuration 'blows up' to
form a $D3$-brane, provided we use the $SU(2)$ ansatz for the transverse scalar fields. This is known as the microscopic description of the brane
polarisation phenomenon.
This funnel solution can be checked from the Abelian, macroscopic, side by considering BIon solutions \cite{bions} on the world-volume.
We wish to know if these funnel solutions can be constructed in general curved backgrounds, and also whether we can
verify the dual picture. This would indeed show that any corrections to the action do not play a role in the large $N$ limit.
Furthermore we expect these funnel solutions to be non-BPS configurations, and so it is of interest to learn whether
the energy will pick up corrections from the symmetrised trace. In flat space the funnel solution for the fuzzy $S^2$ is BPS and it was
shown \cite{costis} that the configuration does not pick up corrections from the symmetrised trace \footnote{Note that the funnel solutions for higher dimensional intersections in flat space are not BPS.}.
Thus one may expect that the BPS condition is protected from any $1/N$ corrections.
We wish to show that this is not quite correct and that it is minimal energy configurations which are protected from such corrections, 
which may not necessarily be BPS, at least for the $D1-D3$ intersection. The explanation of the flat space result is that the BPS condition coincides with the minimum
energy condition.

Another related issue has been the automorphisms of the equations of motion. In \cite{costis} it was shown that
there are dualities between fuzzy funnel solutions and brane dynamics in flat space. The more general case of a curved
background would appear to impose additional constraints on the theory which breaks at least some part of this duality,
just as in the case of adding world-volume electric fields \cite{bhattacharyya}. The automorphisms of the equations
are related to large/small dualities where $r \to 1/r$ in flat space, however we may expect this not to hold in
general curved backgrounds.

We begin with a review of the non-Abelian DBI in a generic curved background, and make some general comments
about the time dependent collapse of a fuzzy sphere with electric fields on the coincident $Dp$-branes.
We then switch our attention to brane intersections in the same background, focusing initially on the $D1-D3$ intersection.
We try to construct the most general funnel solution consistent with the equations of motion and the minimisation of 
energy, before specialising to a few special brane backgrounds. We comment on the automorphisms of the equations of motion
 and the dualities present in curved space before going on to construct the dual Abelian theory of
a $D3$-brane with non trivial magnetic flux on the world-volume. We show that the solutions on the Abelian side are
the same as those on the non-Abelian side. We close with an extension of the work to higher dimensional funnel solutions
before closing with some remarks and possible future directions
\section{Dynamics of Non-Abelian DBI.}
In this section we wish to consider the dynamics of $N$ coincident $Dp$-branes in a curved background,
when there is a homogenous electric field on the world-volume of each of the $N$ branes.
We will begin with type II string theory in ten dimensions, and assume that there is a curved background
generated by some source with $M$ units of flux. The only constraint we will impose on the form of the
background metric is that it is diagonal, with a symmetry group given by $SO(1,q) \times SO(9-q)$
\begin{equation}\label{eq:metric}
ds^2 = -g_{00}dt^2 + g_{xx}dx^a dx^b \delta_{ab} + g_{zz} dz^i dz^j \delta_{ij}
\end{equation}
where $a, b$ run over the $q$ worldvolume directions and $i, j$ are transverse directions to the source. This background could obviously
be generated by a stack of coincident branes, or something more exotic. In this respect this is the generalisation of
the results obtained in to include other background solutions.

Into this background we wish to introduce our $N$ coincident $Dp$-branes, which will have an effective action
given by the Non-Abelian extension of the simple DBI \cite{nonabelianDBI}. The key aspect of this is that we want
to consider these branes as probes of the background geometry, and must therefore ensure that the number of coincident branes
is less than the charge from the background source.
The bosonic part of the action can be written as follows
\begin{equation}\label{eq:action}
S = -\tau_p \int d^{p+1} \zeta STr\left(e^{-\phi}\sqrt{-det(\mathbf{P}[E_{ab}+E_{ai}(Q^{-1}-\delta)^{ij}E_{jb}+\lambda F_{ab}])}\sqrt{det Q^i_j}\right),
\end{equation}
where, as usual, $\mathbf P$ denotes the pullback of the bulk spacetime tensors to each of the brane world-volumes.
In addition we have the following definition for the induced metric $E_{\mu \nu}=G_{\mu \nu}+B_{\mu \nu}$, where
$G_{\mu \nu}, B_{\mu \nu}$ are the bulk metric and Kalb-Ramond two form respectively. However for the remainder of
this note we will set $B=0$ for simplicity. The open string couplings on the world-volume are controlled by the 
inverse of the F-string tension as $\lambda = 2\pi \alpha'$, where $\alpha' = l_s^2$ is the slope of the Regge trajectory
and equal to the square of the string length. The last term in the action is often referred to as the potential,
as in the non-relativistic limit we recover a dimensionally reduced Yang-Mills theory. The full expression for the
matrix $Q$ is given by $Q^i_j = \delta^i_j + i\lambda[\phi^i, \phi^k]E_{kj}$, where the $\phi^i$ are the transverse
coordinates to the $p$-branes world-volume.

Recall that because the $N$ branes are now coincident, the strings that stretched between each brane are now
massless and fill out extra degrees of freedom to enhance the world-volume symmetry from $U(1)^N \to U(N)$. Thus
the deformations of the world-volume, corresponding to excitations of the string ends, must now transform as
$N \times N$ matrices in the adjoint representation of this new $U(N)$ gauge group. The only ambiguity is how to
obtain scalars from matrix valued objects, which is accomplished with the use of the symmetrised trace - denoted by STr.
The prescription for taking this trace is to firstly take the symmetrised average over all orderings of $F_{ab}, D_a \phi^i, 
i[\phi^i, \phi^j]$ before taking the trace. Obviously now that we are dealing with non-Abelian gauge groups we must
use the covariant derivate in the pullback operation.

We are interested in the dynamics of this configuration, and so we will demand that our transverse scalar fields are
time dependent only, namely $\phi = \phi(t)$. Additionally we will begin by using diffeomorphism invariance
to position the branes parallel to the gravitational source, but displaced along one of the transverse directions.
On each of the world volumes we will also turn on an electric field using $F_{0a}=\varepsilon_a$ where $a, b = 1 \ldots p $
are world-volume directions, and we implicitly assume that we take the $A_0=0$ gauge and that the gauge field
commutes with itself. It will often be convenient to write $\varepsilon^2 = \sum_{a} \varepsilon_a \varepsilon^a$ for simplicity.
For the sake of generality we will assume that the gauge field is homogenous on the world-volume
of the coincident branes.
After calculating the determinant, the kinetic part of the action can be seen to reduce to the following form
\begin{equation}
S_{kin} =-\tau_p \int d^{p+1} \zeta STr \left(e^{-\phi}\sqrt{g_{xx}^p g_{00} (1-\lambda^2 g_{zz}g_{00}^{-1}\dot{\phi^i}\dot{\phi^j}\delta_{ij}-\lambda^2 \varepsilon^2g_{xx}^{-1}g_{00}^{-1})}\right),
\end{equation}
where we must still perform the symmetrised trace over the adjoint indices. 
It will be useful for us to make an ansatz for the transverse scalars which reflects the non-Abelian
group structure of the theory, which we can do using the 'simple' group $SU(2)$. This can be accomplished by setting all but three of the transverse scalars to zero, where
we then impose the condition
\begin{equation}
\phi^i = R(t) \alpha^i,
\end{equation}
where the $\alpha^i$ are the generators of the $N \times N$ representation of the $SU(2)$ algebra satisfying the usual commutation relation
\begin{equation}
[\alpha_i, \alpha_j]= 2i\epsilon_{ijk}\alpha_k.
\end{equation}
Strictly speaking this ansatz should be imposed upon the complete equations of motion and not upon the action, however it transpires that the
ansatz is indeed consistent. Upon substitution of our ansatz into the kinetic part of the action written above we find it reduces to
\begin{equation}
S_{kin} =-\tau_p \int d^{p+1} \zeta STr \left(e^{-\phi}\sqrt{g_{xx}^p g_{00} (1-\lambda^2 g_{zz}g_{00}^{-1} \dot{R}^2 \alpha^i \alpha^i-\lambda^2 \varepsilon^2g_{xx}^{-1}g_{00}^{-1})}\right).
\end{equation}
In order to take the trace of this expression we need to Taylor expand the action and then deal with all the symmetrised contributions of the various powers of the generators, however
we can make some headway by taking $N$ to be large implying that all the $1/N$ correction terms are negligible and can be dropped from the action.
This limit is acceptable because we wish to neglect gravitational back reaction. We can accomplish this by decoupling the closed string sector of the theory, namely
sending the factor $g_s \to 0$. However, we also need to keep $g_s N <1$ and fixed whilst taking this limit and therefore we
are forced into taking the large $N$ limit.
We also note that the metric components are generally functions of the transverse coordinates, which implies that
they will be proportional to a trace over the group generators. However, the radial coordinate implicit in the anzatz
is not of the correct dimensionality and thus we are forced to use the physical distance in the metric functions.
The implications for this are potentially far reaching, as we are assuming that the metric (and dilaton) terms are
singlets with respect to the symmetrised trace. Thus we are treating the background as a semi-classical geometry, and
fully expect there to be sub-leading corrections which reflect the quantum nature of the theory. 
With these remarks in mind, and using the definition
of the quadratic Casimir $CI_n= \sum_i \alpha^i \alpha^i = (N^2-1)I_n$, we can pull various terms
through the trace operation and write the full action as follows
\begin{equation}
S= -\tau_p \int d^{p+1}\zeta N g_{xx}^{p/2}g_{00}^{1/2}e^{-\phi}\sqrt{(1-g_{zz}g_{00}^{-1}\lambda^2 C \dot{R}^2-
g_{xx}^{-1}g_{00}^{-1}\lambda^2 \varepsilon^2)(1+4g_{zz}^2\lambda^2CR^4)},
\end{equation}
where we are making the reasonable assumption that the dilaton term is a c-number with respect to the trace operation.
Varying this term with respect to $\dot{R}$ and $\varepsilon^a$ yields the canonical momenta for the radial mode and the displacement field respectively, the latter term being
\begin{equation}
D^a = \frac{\tau_p V_p e^{-\phi} g_{xx}^{p/2}g_{00}^{1/2} \sqrt{1+4\lambda^2CR^4g_{zz}^2}}{\sqrt{1-g_{zz}g_{00}^{-1}\lambda^2 C \dot{R}^2-g_{xx}^{-1}g_{00}^{-1}\lambda^2 \varepsilon^2}} \left(\frac{\lambda^2 \varepsilon^a}{g_{xx}g_{00}}\right),
\end{equation}
where we note that $D^a$ is the electric flux along the $x^a$ direction on each of the world-volumes and is related to the charge of the fundamental string.
As usual, the canonical momenta allows us to construct the Hamiltonian via Legendre transform
\begin{equation}
\mathcal{H} = \frac{\tau_p V_p N e^{-\phi}g_{xx}^{p/2}g_{00}^{1/2}\sqrt{1+4g_{zz}^2\lambda^2 CR^4}}{\sqrt{1-g_{zz}g_{00}^{-1}\lambda^2 C \dot{R}^2-g_{xx}^{-1}g_{00}^{-1}\lambda^2\varepsilon^2}}.
\end{equation}
At this juncture we note that $R$ is not the physical distance of the probe branes from the source, however the two distances are related via the expression
\begin{equation}\label{eq:physical_distance}
r^2 = \frac{\lambda^2}{N}Tr(\phi^i \phi^j \delta_{ij})=\lambda^2C R^2, 
\end{equation}
and so we may write the physical Hamiltonian as follows
\begin{equation}\label{eq:energy}
\mathcal{H}_{phys} = \frac{\tau_p V_p N e^{-\phi}g_{xx}^{p/2}g_{00}^{1/2}}{\sqrt{1-g_{zz}g_{00}^{-1}\dot{r}^2-g_{xx}^{-1}g_{00}^{-1}\lambda^2\varepsilon^2}}
\sqrt{1+\frac{4g_{zz}^2r^4}{\lambda^2 C}},
\end{equation}
or we can write it in the Hamiltonian formalism
\begin{equation}
\mathcal{H}_{phys} = \sqrt{\left(\tau_p V_p N e^{-\phi}g_{xx}^{p/2}g_{00}^{p/2} \right)^2 \left(1+\frac{4r^4g_{zz}^2}{\lambda^2 C} \right) + \frac{g_{00}\Pi^2}{g_{zz}\lambda^2C}+\frac{D^2 g_{xx}g_{00}}{\lambda^2}}
\end{equation}
In the above expressions we have defined $V_p$ as the $p$-dimensional volume element of the branes. Note that when $p=0$, corresponding to coincident $D0$-branes, the electric field contribution vanishes, as it must
since the world-volume cannot support a rank two field strength tensor. In general the Hamiltonian will be conserved, however $\varepsilon^a$ will not. This is because it is the flux that is the conserved
charge on the $D$-brane, and not the gauge field. However because of our homogenous ansatz we find that the electric field is conserved in this instance, and so we may write it as follows
\begin{equation}
\varepsilon^a = \frac{D^a}{\mathcal{\tilde{H}}},
\end{equation}
which shows us that the electric field is conserved and quantised with $D$ units of charge.
\subsection{Minkowski space dynamics.}
We have tried to keep the background space-time as general as possible, however in this section we will consider the dynamics of these branes in the flat space limit. The situation can be described
as follows. We have $N$ coincident $Dp$-branes with three excited transverse scalar fields parameterising a fuzzy two-sphere, the physical radius of which is given by $r$. 
The flat space Hamiltonian can be written simply as
\begin{equation}
\tilde{\mathcal{H}} = \frac{1}{\sqrt{1-\dot{r}^2-\lambda^2\varepsilon^2}} \sqrt{1+\frac{4r^4}{\lambda^2C}},
\end{equation}
where we introduce the simplifying notation $\tilde{\mathcal{H}} = \mathcal{H}/(\tau_p V_p N)$, and note that $\dot{r}$ corresponds to the velocity of the collapsing fuzzy sphere.
Furthermore with this definition of the Hamiltonian we lose all dependence on the dimensionality of the probe $Dp$-branes. Thus in the Minkowski limit all the
$p$-branes yield the same equations, another example of $p$-brane democracy.
As is usual with this type of problem it will be more convenient for us to employ the use of dimensionless variables \cite{costis, rst}. By making the following definitions
\begin{equation}\label{eq:dimensionless}
z= \sqrt{\frac{2}{\lambda \sqrt{C}}}r, \hspace{0.5cm} \tau = \sqrt{\frac{2}{\lambda \sqrt{C}}}t, \hspace{0.5cm} e=\lambda \varepsilon,
\end{equation}
the Hamiltonian and effective potential can be written as follows
\begin{eqnarray}
\tilde{\mathcal{H}} &=& \sqrt{\frac{1+z^4}{1-\dot{z}^2-e^2}} \nonumber \\
V_{eff} &=&  \sqrt{\frac{1+z^4}{1-e^2}}.
\end{eqnarray}
The electric field must satisfy the usual constraint $e^2 \le 1$ in order for the theory to remain valid. The other constraint can be seen to be $1 \ge \dot{z}^2+e^2$, which implies
that the velocity of the collapse is reduced by a factor $\sqrt{1-e^2}$, which is less than the speed of light
For an arbitrary field strength we see that the fuzzy sphere will tend to collapse down to zero size as expected.

Our Hamiltonian has no explicit time dependence and is therefore a conserved charge which will allow us to obtain a solution to the equation of motion. We choose the initial
conditions $\dot{z}(0)=0$ and $z(0)=z_0$ to indicate an initially static configuration at some arbitrary distance $z_0$. By integrating the equation of motion and using the many
properties of Jacobi Elliptic functions, we arrive at the solution
\begin{equation}\label{eq:flat_solution}
z(\tau) = \pm z_0 {\rm{JacobiCN}} \left\lbrack \frac{\sqrt{2(1-e^2)}\tau z_0}{\sqrt{1+z_0^4}}, \frac{1}{\sqrt{2}} \right\rbrack.
\end{equation}
Note that $z_0$ corresponds to the initial radius of the fuzzy sphere
(in dimensionless variables). Taking the positive sign initially, one sees that as time evolves the fuzzy sphere
collapses. The speed of the collapse is dependent upon the strength of the electric field, because an increasing
field implies that the branes move more slowly. The physical interpretation of this is that the extra flux on the
world-volume acts as extra 'mass', which acts to reduce the velocity. If there is a critical electric field
which saturates the bound $e^2=1$ then the fuzzy sphere will be static for all time. This is different to the 
result obtained when considering the dynamics without gauge fields, which always implied collapsing solutions - at least
to leading order in $1/N$. Eventually the sphere reaches zero size,
however the periodic nature of the solution appears to imply re-expansion into a region of negative $z$. This 
is due to the ambiguity in taking the positive sign for the physical radius in (\ref{eq:physical_distance}) \cite{costis}.
A similar remark applies when taking the minus sign in the above solution. Note that in both cases, it is the $R^2$ term
that appears in the DBI action and therefore no potential for discontinuities when we use the different sign choices
for the physical radius.

The zeros of the elliptic function occur when the amplitude equals $K(k)$, where $K$ is the complete elliptic integral of the first kind.
This allows us to calculate the collapse time $t_{*}$ for the fuzzy sphere to be
\begin{equation}
\tau_{*}= \sqrt{\frac{1+z_0^4}{2(1-e^2)}}\frac{1}{z_0} K\left(\frac{1}{\sqrt{2}}\right),
\end{equation}
which agrees with our intuitive notion that by increasing the electric field, the collapse takes longer to occur.
\subsection{1/N Corrections in Minkowski space.}
In this section we will investigate the corrections to the theory arising from the symmetrised trace prescription. These corrections were
first derived in \cite{rst}, and we refer the interested reader to that paper for more details. In flat space it was emphasised that as the fuzzy sphere collapses its velocity
approaches the speed of light, and therefore higher order terms in $1/N$ ought to become important in order to fully describe the dynamics. This is due to the fact that
the energy will increase as the velocity increases. It has been argued that these corrections are all zero for a BPS object, however we suspect that this is only true for
flat space configurations where the requirement of minimal energy is satisfied by an object being BPS.
However the presence of an electric field on the brane world-volumes reduces the velocity of the collapse by the factor $\sqrt{1-e^2}$ and thus the leading order Lagrangian
may remain valid - although there are difficulties associated with near critical electric fields and the DBI \cite{dorn}. In curved space the gravitational red shift appears
to reduce the velocity of the fuzzy sphere to sub-luminal speeds, however there was found to be no turning point solution in the static potential and therefore
no formation of non-Abelian bound states (with the exception of $D0$-branes in the $D6$-brane background.)

The important result from  is that the corrections to the Lagrangian an be written as a series expansion in powers of $C$, thus our Hamiltonian can be shown to be the $0th$ order in this expansion
\begin{equation}\label{eq:str_corrections}
\tilde{\mathcal{H}} = \left(1-\frac{2C}{3}\frac{\partial^2}{\partial C^2} + \frac{14}{45}C^2 \frac{\partial^4}{\partial C^4} + \ldots \right)\tilde{\mathcal{H}_0}.
\end{equation}
It will be convenient in what follows to return the original action for a flat background, and define the following dimensionless parameters
\begin{eqnarray}
\tilde{r}^4 &=& 4\lambda^2 C R^4\\
\tilde{s}^2 &=& \lambda^2 C \dot{R}^2 \nonumber \\
e^2 &=& \lambda^2 \varepsilon^2 \nonumber
\end{eqnarray}
where the last expression has already been introduced in the previous section. The first two equations can be regarded as defining complex parameters, constrained by a single equation - namely
the conservation of energy, and can be regarded as a 'radial' variable and a 'velocity' variable respectively.
In terms of these complex parameters we can define the Hamiltonian to be 
\begin{equation}
\tilde{\mathcal{H}} = \sqrt{\frac{1+\tilde{r}^4}{1-\tilde{s}^2-e^2}} = U \gamma,
\end{equation}
where $U$ can be regarded as a position dependent mass term, whilst $\gamma$ is the modified relativistic factor as usual. Position dependent masses arise often in physics, in semiconductors for example.
If we now apply the leading order symmetrised trace correction to this form of the Hamiltonian we obtain the following solution
\begin{equation}
\tilde{\mathcal{H}_1} = U\gamma - \frac{\gamma}{6CU^3}\left[3U^4\gamma^4(1-e^2)^2-4U^4\gamma^2(1-e^2)-2U^2\gamma^2(1-e^2)+4U^2-1\right],
\end{equation}
which represents the $1/N$ correction to the Hamiltonian in flat space. The first thing to note is that when there is a critical (or near critical) electric field, the corrected
Hamiltonian reduces to
\begin{equation}
\tilde{\mathcal{H}_1} \sim \frac{U}{i \tilde{s}} \left(1-\frac{(4U^2-1)}{6CU^4} \right),
\end{equation}
which is clearly imaginary and therefore does not correspond to a physical solution. We can avoid this problem by rotating the background metric to a Euclidean signature
and studying the effects of over-critical electric fields, however we will not do that in this instance \footnote{We refer the interested reader to the recent work \cite{dorn} for more information.}.

More generally we will have an arbitrary non-critical electric field, however we can still learn about the physical interpretation of the energy corrections.
We first consider the static solution, i.e zero velocity, in which case the Hamiltonian becomes
\begin{equation}
\mathcal{H}_1 = \frac{U}{\sqrt{1-e^2}}\left(1-\frac{(2U^2-U^4-1)}{6CU^4} \right).
\end{equation}
The correction terms will be non-zero except for when we choose $U=1$, or when $U\to \infty$ corresponding to large radius. In this latter limit we would expect the geometry to
resemble the classical geometry of the two-sphere. It should be noted that there is no value of $r$ for which the energy will vanish.
If we now consider the case where $R \to 0$, the energy reduces to
\begin{equation}
\mathcal{H}_{1} = \gamma \left(1-\frac{(\gamma^2(1-e^2)^2-2\gamma^2(1-e^2)+1)}{2C} \right).
\end{equation}
The correction term will be minimised by sending $\tilde s  \to 0$, however it can be seen that the  Hamiltonian itself will vanish if the velocity term satisfies
\begin{equation}
\tilde s^2 = (1-e^2) \left(1-\frac{1}{1 \pm \sqrt{2}N} \right) \sim (1-e^2)
\end{equation}
where we have explicitly taken the large $N$ limit. Note that when the electric field is zero this condition reduces to $\tilde s^2 =1$, implying that the
branes are moving at the speed of light. Therefore in general we see that increasing the strength of the electric field reduces the velocity of the branes, as expected, and 
therefore can reduce the energy of the configuration when it is located at the origin.
\subsection{Curved space dynamics.}
The dynamics of the fuzzy sphere in a curved background are generally non-trivial due to the 
additional dependence of the metric components, and dilaton, upon the embedding coordinates. Thus we can only
obtain exact solutions by specifying the form of the background. We repeat the physical Hamiltonian here for 
convenience.
\begin{equation}
\tilde{\mathcal{H}} = \frac{e^{-\phi}g_{xx}^{p/2}g_{00}^{1/2}}{\sqrt{1-g_{zz}g_{00}^{-1} \dot{r}^2-g_{xx}^{-1}g_{00}^{-1}\lambda^2 \varepsilon^2}}
\sqrt{1+\frac{4g_{zz}^2r^4}{\lambda^2 C}}, \nonumber
\end{equation}
which allows us to define the static potential as follows
\begin{equation}
V = \frac{e^{-\phi}g_{xx}^{p/2}g_{00}^{1/2}}{\sqrt{1-g_{xx}^{-1}g_{00}^{-1}\lambda^2 \varepsilon^2}}\sqrt{1+\frac{4g_{zz}^2 r^4}{\lambda^2 C}}
\end{equation}
The unknown dependence of the metric components upon the physical radius prevents us from determining the general
behaviour of the fuzzy sphere in this background. However we can see that the maximum value for the electric field
will be a function of the transverse variables and therefore the radius of the fuzzy sphere. 
The general solution for the maximal field value can be seen to be
\begin{equation}
\varepsilon_{max} \le \frac{\sqrt{g_{00}g_{xx}}}{\lambda}.
\end{equation}
In our analysis we will assume that the electric field does not saturate this bound in order to keep the action finite and 
real. There has been extensive work on overcritical fields on $D$-branes, but this will not be relevant here.
Using the conservation of the Hamiltonian we find the general expression for the velocity of the collapsing fuzzy sphere
\begin{equation}
\dot{r} = \frac{\delta \mathcal{H}}{\delta \Pi} = \left( \frac{\Pi g_{00}}{\mathcal{H} g_{zz} \lambda^2 C}\right).
\end{equation}
Now for general supergravity solutions we expect the metric components corresponding to the $SO(1,q)$ directions to correspond
to either flat, or decreasing monotonic functions of the physical radius. Conversely we would anticipate that the $g_{zz}$ functions are either
flat, or increasing monotonic functions of $r$ - becoming singular when we reach zero radius. Therefore the general expression for the velocity suggests that
it is a decreasing function of the physical radius regardless of the specific values of the ratio of $\Pi / \mathcal{H}$, provided that it is finite.
The implication for this is that the sphere would take an infinite amount of time to collapse to zero size, neglecting any open string effects at short distances.
This 'braking' behaviour is in contrast to what happens in flat space, where the fuzzy sphere collapses at an ever increasing velocity. However this
is in a gravitational background and we expect the velocity term to be red shifted by the factor $g_{00}/g_{zz}$, thus by switching to proper time variables we 
would find that the collapse occurs in finite time.

The acceleration of the sphere turns out to be
\begin{equation}
\ddot{r} = \frac{\Pi \dot{r}}{\mathcal{H}g_{zz}\lambda^2 C}\left(g_{00}^{\prime}-\frac{g_{zz}^{\prime}}{g_{zz}} \right),
\end{equation}
where primes denote derivatives with respect to the physical radius. The equation can be seen to be zero in three cases, firstly when $\dot{r}$ is zero which is the trivial solution as the sphere is static. 
Secondly when $g_{zz} \to \infty$ which implies
that we must take $r \to 0$ and so the effective action breaks down, and finally when we have the case $g_{00} = \rm{ln}(g_{zz})$.
Provided the derivatives of the metric function are continuous, we see that the acceleration will never become singular and so we would expect the DBI to provide a reasonable description of the dynamics
of the coincident branes.

At this point it is useful to consider some concrete examples of non-trivial backgrounds in order to fully understand the
dynamical collapse of the fuzzy sphere.
\subsubsection{$Dq$-brane background}
The supergravity solution for a background generated by coincident $Dq$-branes is given by the following
\begin{eqnarray}\label{eq:dbranesoln}
ds^2 &=& H^{-1/2} \eta_{\mu \nu} dx^{\mu} dx^{\nu} + H^{1/2} dz^{a}dz^{b} \delta_{ab} \\
e^{-\phi} &=& H^{(q-3)/4} \nonumber \\
H &=& 1 + \frac{k_q}{r^{7-q}} = 1+ \frac{(2\sqrt{\pi})^{5-q}M g_s \Gamma(\frac{7-q}{2})l_s^{7-q}}{r^{7-q}}, \nonumber
\end{eqnarray}
where $r$ is the physical distance from the source branes, and also the radius of our fuzzy sphere. As usual $g_s, l_s$ are the string coupling and string length respectively.
It must be remembered that $q$ even corresponds to type IIA string theory, whilst $q$ odd corresponds to type IIB string theory. As we are considering the general case
of $Dp$-branes in a $Dq$-brane background, we can neglect the RR couplings arising from the background branes.
Upon identification of the various metric components we write the conserved Hamiltonian as follows
\begin{equation}
\tilde{\mathcal{H}} = \frac{H^{(q-p-4)/4}}{\sqrt{1-H\dot{r}^2-H\lambda^2 \varepsilon^2}}\sqrt{1+\frac{4Hr^4}{\lambda^2C}}
\end{equation}
and the expression for the static potential becomes
\begin{equation}
V_{eff} = \frac{H^{(q-p-4)/4}}{\sqrt{1-H\lambda^2 \varepsilon^2}}\sqrt{1+\frac{4Hr^4}{\lambda^2 C}}
\end{equation}
The last expression tells us that the electric field can diverge as the radius of the fuzzy sphere collapses.
\subsubsection{$NS$5-brane background.}
The $NS$5-brane background has been extensively researched of late, as it is a simple non-trivial solitonic background
which also has links to little string theory. The supergravity solutions for $M-NS$5 branes are shown below, note that 
they are invariant under T-duality because the harmonic function only couples to the transverse components of the metric
\begin{eqnarray}\label{eq:ns5soln}
ds^2 &=& \eta_{\mu \nu} dx^{\mu}dx^{\nu} + H dz^{a}dz^{b} \delta_{ab} \\
e^{-\phi} &=& H^{-1/2} \nonumber \\
H &=& 1 + \frac{Ml_s^2}{r^2}. \nonumber
\end{eqnarray}
The expression of interest for us is the static potential, which can be seen to reduce to
\begin{equation}
V_{eff} = \frac{1}{\sqrt{H}\sqrt{1-\lambda^2 \varepsilon^2}}\sqrt{1+\frac{4H^2r^4}{\lambda^2 C}},
\end{equation}
implying that the maximal electric field bound is $\varepsilon_{max} \le \lambda^{-1}$.
It is straight-forward to see that there is no turning point for the potential, except when we take $r$ to be large
which corresponds to the global maximum. The implication is that there is no radius at which the fuzzy sphere may stabilise at,
and therefore nothing to halt the progress of the probe branes toward the five-branes even with the inclusion of an electric field.
The analysis of this solution was detailed in \cite{thomas}, and we refer the interested reader there for more information.
\subsubsection{$F$-string background.}
We can also consider the background sourced by $M$ fundamental strings, where for consistency we should limit
the dimensionality of the probe branes to $p \le 1$ in order to fully justify our assumption about neglecting
backreaction effects, the resulting configuration is a bound state of fundamental strings and $D$-strings more commonly referred to as 
as $(p,q)$-string. The supergravity background solution is
\begin{eqnarray}\label{eq:fstringsoln}
ds^2 &=& H^{-1} \eta_{\mu \nu} + dz^{a}dz^{b} \delta_{ab} \\
e^{-\phi} &=& H^{1/2} \nonumber \\
H &=& 1 + \frac{2^5 \pi^2 g_s^2 l_s^6 M}{r^6} \nonumber,
\end{eqnarray}
where now $\mu, \nu$ run over one temporal and one spatial dimension. The static potential for the bound state can be written
\begin{equation}
V_{eff} = \frac{1}{H}\sqrt{H\tilde \tau_1^2 \left(1+\frac{4r^4}{\lambda^2 C} \right) + \frac{\Pi^2H}{\lambda^2 C}+\frac{D^2}{\lambda^2}},
\end{equation}
where we have rescaled the $D1$-brane tension such that $\tilde \tau_1 = \tau_1 V_1 N$. Thus we effectively have a $(D, N)$-string bound state.
We are at liberty to consider various limits of the potential, however the general behaviour is that it is always a monotonically decreasing function of the radius.
For the $D$-string dominated solution we find the Hamiltonian scales like the tension of the string on a fuzzy sphere, namely
\begin{equation}
\mathcal{H} \sim \sqrt{\frac{\tilde \tau_1^2}{H}\left(1+\frac{4r^4}{\lambda^2 C} \right)}.
\end{equation}
Conversely, taking the $F$-string dominated solution we find that the Hamiltonian scales with the displacement field
\begin{equation}
\mathcal{H} \sim \frac{|D|}{H \lambda},
\end{equation}
which shows that both configurations will be gravitationally attracted toward the $F$-string background as this is the lowest energy state.
The background string coupling tends to zero with the physical radius of the fuzzy sphere which means that our world-volume description can be
trusted to very late times. As the strings move closer together we expect the formation of a new $(D+M, N)$-string bound state. The binding energy of which can
be shown to be of the form
\begin{equation}
E_{\rm bind} \sim \sqrt{\tilde \tau_1 ^2 + (D+M)^2.}
\end{equation}
This result mimics the behaviour in the Abelian theory where it was shown that the condition $g_s \to 0$ with $g_s N >>1$ prevented the emission of
closed string states and as such could be regarded as a semi-classical field theory.
We close with a remark about the electric field in this instance. In the large radius limit we find that the displacement field can be well approximated by
\begin{equation}
D \sim \frac{2 \tilde \tau_1 \lambda \varepsilon_{\infty}r^2}{\sqrt{C}\sqrt{1-\lambda^2\varepsilon_{\infty}^2}},
\end{equation}
where $\varepsilon_{\infty}$ reflects the strength of the field at large distances where the harmonic function is approximately unity. As the sphere collapses,
the electric field is driven to its critical value, resulting in an increase in the displacement field. This behaviour can be seen via the expression
\begin{equation}
D^a= \mathcal{H}\lambda^2 H^2 \varepsilon^a,
\end{equation}
where the right hand side naturally becomes large as the radius shrinks. This tells us that at exactly at the threshold point of the bound state, the electric field reaches its critical
value and the string becomes tensionless. A more detailed analysis with the inclusion of angular momentum modes would tell us a great deal about the formation of this bound
state. We have also assumed here that the closed string modes will be suppressed, however a more detailed investigation would be useful as the supergravity constraints
impose the strong condition $M >> N$ if we are to neglect back reaction. This is potentially useful in the investigation of cosmic superstring networks \cite{cosmicstrings}.

\subsection{1/N corrections in Curved space.}
As in the flat space case we can consider higher order corrections to the energy in powers of $1/N$ coming from the application of the symmetrised trace.
We will find it convenient to define the following variables
\begin{equation}
\alpha = e^{-\phi}g_{xx}^{p/2}g_{00}^{1/2} \hspace{0.5cm} \beta=\sqrt{1+4g_{zz}^2R^4\lambda^2 C} \hspace{0.5cm} \gamma= (1-e^2-g_{zz}g_{00}\dot{R}^2\lambda^2C)^{-1/2},
\end{equation}
where $e^2 = g_{xx}^{-1}g_{00}^{-1}\lambda^2 \varepsilon^2$ and therefore the Hamiltonian reduces to $\tilde{\mathcal{H}} = \alpha \beta \gamma$. We know that this 
energy is the zeroth order expansion in powers of $1/N$ and using (\ref{eq:str_corrections}) we find that the first order Hamiltonian is remarkably similar to that constructed in the flat space
instance
\begin{equation}
\tilde{\mathcal{H}_1} = \alpha \beta \gamma - \frac{\alpha \gamma}{6C\beta^3}\left(3\beta^4\gamma^2(1-e^2)-4\beta^4\gamma^2(1-e^2)-2\beta^2\gamma^2(1-e^2)+4\beta^2-1 \right)
\end{equation}
In deriving this expression we are explicitly assuming that the metric components are unaffected by the symmetrised trace prescription. Note that the energy to all orders will depend on
the $\alpha$ factor, and therefore when this is zero the energy of the configuration will be zero. As we have argued, in general the metric functions $g_{00}$ and $g_{xx}$ are decreasing functions
of $r$, so that they vanish when $r \to 0$ so we may expect that the energy will always tend to zero. However we have no way of knowing the general behaviour of the dilaton term with respect to
the radial distance. What is clear is that minimising $\alpha$ is equivalent to minimising the energy.
We begin by considering the case of zero electric field, we choose to set $\varepsilon = 0$ rather than taking the limit of the metric components to zero as this will imply that $\alpha$ and therefore
the energy is zero. We further wish to consider the static case, with the branes at an arbitrary distance away from the source. This reduces the Hamiltonian to the following form
\begin{equation}
\tilde{\mathcal{H}_1} = \alpha\beta -\frac{\alpha}{6C\beta^3}(2\beta^2-\beta^4-1).
\end{equation}
There will be no correction terms when $\beta = 1$, which corresponds to the two cases $R \to 0$ or $g_{zz}^2 \to 0$. The first of these implies that $r\to 0$ and so the branes will be on
top of the sources where we expect the DBI to break down. The second case corresponds to sending $r\to \infty$ because the metric component is generally an increasing function as $r\to 0$. This latter
limit is unphysical in our situation, and so we see that the sphere energetically favours collapse from a static position. There will also not be any corrections as $\beta \to \infty$, which implies
that either $r \to \infty, 0$ leading to the same remarks as above.

We now insist on keeping the electric field turned on, although the modification to the Hamiltonian in the static limit is very similar to the zero field case. The solution reduces to
\begin{equation}
\tilde{\mathcal{H}_1} = \frac{\alpha \beta}{\sqrt{1-e^2}} \left(1-\frac{(2\beta^2-\beta^4-1)}{6C\beta^4} \right).
\end{equation}
In this case the correction terms will only vanish as $\beta \to \infty$, which corresponds to the case of infinite energy for the fuzzy sphere.
Thus for finite electric field we see that the solution will still collapse toward zero size provided that the dilaton term does not blow up in the small $r$ limit. In fact this is what distinguishes
the $D6-D0$-brane system from the others as this is precisely where the dilaton term becomes large as the same time that the other metric components are going to zero. The resultant energy profile
is not monotonic but yields a stable minimum in which a bound state can form \cite{hyakutake, thomas}.
The case of critical, or almost critical field, is similar to the flat space scenario, where the energy becomes imaginary.
\section{Brane Intersections in Curved Space.}
Thus far our our analysis has dealt with parallel brane configurations, however this is not the only place non-commutative
geometry enters in string theory as we can also consider intersecting branes. The simplest intersections have been
investigated in a series of papers, where $N D1$-branes intersect with either $D3$, $D5$ or $D7$-branes in flat space-time \cite{constable, cook, koch, costis, bhattacharyya}.
There are two dual world-volume descriptions of the intersection. The first is from the higher dimensional brane
viewpoint, where the $D1$-brane is realised as an Abelian BIon spike solution in a transverse direction. 
For the $D3$ scenario it is necessary to turn on a homogenous magnetic field on the brane, since the $D$-string acts as a 
magnetic monopole solution. The $D5$ world-volume description is more complicated because we have a non vanishing
second Chern class. The dual picture is from the non-Abelian viewpoint of the $D$-strings, which can be seen to blow
up into the higher dimensional branes when we use non-commutative co-ordinates. 

In this section we will investigate the $D1-D3$ intersection in the generic, static curved background labelled by the 
metric solution (\ref{eq:metric}), with the inclusion of a constant electric field along the string world-volume. As such, in contrast to Section 2, we will 
not consider $\varepsilon $ as a dynamical degree of freedom.
In fact the addition of a constant electric field turns the $D$-string
into a $(p,q)$-string as the electric field can be interpreted as the dissolving of the fundamental string degrees of freedom into the world-volume.
We will assume that the string is oriented in the $X_0 - X_9$ plane, where we will take $X_0 = t$ and $X_9 = \sigma$
to parameterise the embedding coordinates.
We will also take the gauge $A_0 = 0$ and assume that the gauge field commutes with the transverse scalars. The kinetic part 
of the action reduces to the following expression
\begin{equation}
S=-\tau_1 \int d^2 \sigma STr\left(e^{-\phi}\sqrt{g_{00}g_{zz}(1-\lambda^2g_{xx}g_{00}^{-1}\dot{\phi_a}\dot{\phi_a} +\lambda^2g_{xx}g_{zz}^{-1}\phi_a^{\prime} \phi_a^{\prime}\-\lambda^2\varepsilon^2 g_{00}^{-1}g_{zz}^{-1})} \right),
\end{equation}
where a dot denotes derivatives with respect to time, and primes are derivatives with respect to $\sigma$. In the above we use the standard notation of representing the matrix-valued world volume scalar fields as $\phi_a$ which are not to be confused with the dilaton field  $ \phi $. As in \cite{constable}  we
simplify our analysis by only considering fluctuations of the $D$-strings perpendicular to their world sheet that are also parallel to the world volume of the source branes.
 As such we look to employ the $SU(2)$ ansatz 
\begin{equation}
\phi^a = R(t, \sigma) \alpha^i, \hspace{1cm} a= 1, 2, 3,
\end{equation}
where the $\alpha^i$ again are the generators of the algebra and $a=1,2,3$ label coordinates parallel to the source branes.
Inserting the ansatz into the full action, and taking the large
$N$ limit produces the following 
\begin{eqnarray}
S=-\tau_1 \int d^2 \sigma &N&e^{-\phi}\sqrt{(g_{00}g_{zz})(1-\lambda^2Cg_{xx}g_{00}^{-1}\dot{R}^2+\lambda^2Cg_{xx}g_{zz}^{-1} R^{\prime 2} - \lambda^2 \varepsilon^2 g_{00}^{-1}g_{zz}^{-1})} \nonumber \\
& & \sqrt{(1+4\lambda^2CR^4 g_{xx}^2)},
\end{eqnarray}
where we have neglected higher order corrections to the DBI, and also ignored any potential Chern-Simons term which may arise from the background source.

The metric components are typically functions of the $9-q$ transverse coordinates to the  source branes. By our simplification above, we 
can consistently set the transverse coordinates to zero with the exception of $x_9 = \sigma$, and thus all the metric components
are now explicit functions of $\sigma$. We will also assume that any dilaton term is purely a function of $\sigma$ in order to simplify our analysis.
In most of what follows we will only consider the near horizon approximation, however we will occasionally make reference to the 
Minkowski limit.

In what follows we shall be interested in either the time dependent solution or the spatial solution. It will be the
latter that defines the fuzzy funnel solution. In any case the diagonal components of the energy-momentum tensor
for the above action can be written as follows
\begin{eqnarray}
T_{00} &=& \frac{ e^{-\phi}\sqrt{g_{00}g_{zz}(1+4\lambda^2 C R^4 g_{xx}^2)}(1+\lambda^2 C R^{\prime 2}g_{xx}g_{zz}^{-1}-g_{00}^{-1}g_{zz}^{-1}\lambda^2 \varepsilon^2)}
{\sqrt{1-\lambda^2 C \dot{R}^2 g_{xx}g_{00}^{-1}+\lambda^2 C R^{\prime 2} g_{xx}g_{zz}^{-1}-g_{00}^{-1}g_{zz}^{-1}\lambda^2\varepsilon^2}} \nonumber \\
T_{\sigma \sigma} &=& \frac{ e^{-\phi}\sqrt{g_{00}g_{zz}(1+4\lambda^2 C R^4 g_{xx}^2)}(1-\lambda^2 C \dot{R}^2 g_{xx}g_{00}^{-1}-g_{00}^{-1}g_{zz}^{-1}\lambda^2 \varepsilon^2)}
{\sqrt{1-\lambda^2 C \dot{R}^2 g_{xx}g_{00}^{-1}+\lambda^2 C R^{\prime 2} g_{xx}g_{zz}^{-1}-g_{00}^{-1}g_{zz}^{-1}\lambda^2\varepsilon^2}}, 
\end{eqnarray}
where we have explicitly divided out each term by the factor $\tau_1 N V_1$ which is independent of any space-time
coordinates and will not affect the equations of motion. We must now consider the static and dynamical cases separately if we wish to find simple solutions to the equations of motion.
\subsection{Funnel solutions.}

We can now attempt to find solutions by specifying the background explicitly. We know that in flat Minkowski space the solutions
correspond to funnels, where the lower-dimensional branes blow up into a solitary $D3$-brane. We may expect these funnel type solutions to
occur in curved space as well, however the form of the solution will be different. Firstly consider a stack of $Dq$-branes, which have the
following supergravity solution
\begin{equation}
ds^2 = H^{-1/2}\eta_{\mu \nu}dx^{\mu} dx^{\nu} + H^{1/2}dx^{i}dx^{j} \delta_{ij}, \hspace{0.5cm} e^{-\phi}=H^{(q-3)/4},
\end{equation}
where $\mu, \nu$ are world-volume directions and $i, j$ are transverse directions. The warp-factor $H$ is a harmonic function in the 
transverse directions, which since we are only considering fluctuations of the $D$-string parallel to the $Dq$ world volume
implies they are only dependent on $\sum_{i=9-q}^{9} (x^i)^2 = \sigma^2 $
 and we assume $q = 1, 3, 5$ only because we are looking at type IIB string theory. The equation of motion
can be satisfied by the following expression
\begin{equation}\label{eq:funnel1}
R^{\prime 2} = \frac{1}{\lambda^2 C H^{-1}}\left(H^{(q-3)/2}\lbrace1+4\lambda^2 C R^4 H^{-1} \rbrace (1-\lambda^2 \varepsilon^2)^2-(1-\lambda^2 \varepsilon^2) \right).
\end{equation}
Note that for critical electric fields the RHS of the expression vanishes which implies that $R=$constant and therefore no funnel solution
regardless of the background. For near critical fields, the solution is approximately constant until we reach the point where $R$ diverges.
Thus the general behaviour is that increasing the strength of the gauge field forces the funnel to alter its shape. The stronger the field, the wider the funnel and the larger the fuzzy sphere radius.
Temporarily setting the electric field to zero brings us back to the $D$-string solution, and the equation of motion reduces as follows
\begin{equation}
R^{\prime 2} = \frac{1}{\lambda^2 C H^{-1}}\left(H^{(q-3)/2}\lbrace1+4\lambda^2 C R^4 H^{-1} \rbrace -1 \right),
\end{equation}
which can be seen to be trivially solved when $q=3$ since the eom reduces to $R^{\prime}=2R^2$ and we recover the
funnel solution \footnote{Note that this is also the BPS condition in flat space, however the $D3$-brane will
also be supersymmetric in the $D3$-brane background and so this is also the BPS condition in this instance.}
\begin{equation}
R(\sigma) = \frac{-1}{2(\sigma-\sigma_0)}.
\end{equation}
The radius of the funnel diverges at $\sigma = \sigma_0$ where the $D$-strings blow up into a $D3$-brane. Note that the minus sign indicates this is a $D3$-brane and not 
a $\bar{D3}$-brane, since the latter will be unstable in the background. In fact the harmonic function drops out of the equations implying the funnel solution 
is insensitive to the curved background. This is due to the vanishing dilaton term. If we insist on the inclusion of the electric field in the
$D3$-brane solution then we can shift variables in the integration to obtain a solution, which is a simple deformation of the standard funnel as we would anticipate \cite{cook}
\begin{equation}
R(\sigma) \sim \frac{-1}{2\sqrt{1-\lambda^2 \varepsilon^2}(\sigma-\sigma_0)}.
\end{equation}
The effect of increasing the electric field is to force the funnel to open up more at smaller values of $\sigma$. In fact for near critical fields we expect the 
funnel to diverge before the point $\sigma_0$, implying that the $D3$-brane is located at a different position to the case of zero field.
The structure of the equation of motion prohibits us from finding an exact solution in the $D5$ and $D1$-backgrounds.

We can also look at the $NS$5-brane background, where the supergravity solutions are given in (\ref{eq:ns5soln}). The solution
with zero electric field can be parameterised by $R^{\prime} = 2R^2\sqrt{H}$,  with $H(\sigma) $ given by (\ref{eq:ns5soln}) with $r^2 = \sigma^2 $.
In the first instance, if we look in the throat approximation (ie dropping the factor of unity in $H$)  we find the funnel solution
\begin{equation}
R(\sigma) = \frac{-1}{2\sqrt{Ml_s^2} \rm{ln}(\sigma/\sigma_0)}.
\end{equation}
Here we have selected the cut-off distance $\sigma_0$ to represent the location of the $D3$-brane in the transverse
space. Because the dilaton term tends to blow up as we approach the fivebranes, we must worry that our solution (being
weakly coupled to neglect backreaction) may not be valid deep in the throat geometry. Therefore this solution
can be trusted when the curvature of the bulk geometry is relatively small.
Interestingly we see the funnel solution is invariant (up to a sign) under $\sigma \to 1/\sigma$, which is related 

to the large/small duality problem \cite{costis}
and standard $T$-duality solutions in type II string theory. 
The change in sign reflects the change in orientation of the $D3$-brane, however as both $D$ and $\bar{D}$-branes
are unstable in the fivebrane background the minus sign is technically irrelevant.
It may be possible to probe further
into the throat using the corrections from the symmetrised trace. The idea would be to use the fact that $g_s N$ is
constant, but take a slightly larger value for the string coupling. In order to compensate for this we must
reduce the number of $D$-strings and therefore extra $1/N$ terms will become important. Using the technology developed
in \cite{rst} we can calculate these corrections and check to see how the funnel solution is modified.
We must also recall that a $D$-brane preserves a different half of the supersymmetry algebra than the fivebranes, therefore
the supersymmetry will be explicitly broken (or at least non-linearly realised).

We can extend our solution above to the case where we keep the full expression for $H$. This yields  
interpolating solution between the throat solution and Minkowski space, given by
\begin{equation}\label{eq:nonabside}
R(\sigma) = \frac{\mp 1}{2\left(\sqrt{Ml_s^2 + \sigma^2}-\sqrt{M l_s^2+ \sigma_0^2}+\sqrt{Ml_s^2} \rm{ln} \left\lbrace \frac{\sigma[\sqrt{Ml_s^2}+\sqrt{Ml_s^2+\sigma_0^2}]}{\sigma_0[\sqrt{Ml_s^2}+\sqrt{Ml_s^2 + \sigma^2}]} \right\rbrace \right)}
\end{equation}
which can be seen to yield the two asymptotic solutions when we take the appropriate limit. This solution is particularly interesting because of the cut-off imposed in the integral. 
On one side of the $D3$-brane we have a semi-infinite string solution (solution with +sign in (\ref{eq:nonabside}) whilst on the other (- sign choice) we have a string of finite length. In the throat approximation we can relate the two solutions through a $\sigma \to 1/\sigma $
duality. The finite length of the string implies that the energy of the solution is finite. This differs dramatically from the Minkowksi space solution where the energy will be infinite as the string is of 
infinite length. The profile of the solution therefore relates a finite energy configuration to an infinite energy configuration. This behaviour may well have an interesting analogue in the Abelian world-volume
theory.

The corresponding funnel solution in the background of fundamental strings (\ref{eq:fstringsoln}) can be obtained from the following expression
$R^{\prime 2} = 4R^4/H$, which gives, in the throat approximation,
\begin{equation}
R(\sigma) = \frac{-2\sqrt{k}}{\sigma^4-\sigma_0^4},
\end{equation}
obviously diverging strongly in the limit that $\sigma \to \sigma_0$. 
Of course there are many other kinds of backgrounds that we are free to consider. As an example we could look at the static Maki-Shiraishi solutions
corresponding to a static black hole geometry \cite{maki}. In this case we see that $R(\sigma) \propto \sigma^{5/2}-\sigma_0^{5/2}$ which implies that the funnel only diverges
at large values of $\sigma$, very far from the event horizon of the black hole. This class of metrics also allows for time dependent solutions, corresponding to gases
of $D0$-branes and may play an important role in the study of matrix cosmology \cite{gibbons, craps}.

Finally, we note that even though it is difficult to obtain an analytic solution of the funnel profile in 
$Dq$-brane (for $q\neq 3)$- backgrounds progress can be made in the large 
$R$ approximation. In this case we find from (\ref{eq:funnel1}) 
\begin{equation}\label{eq:funnel2}
R^{\prime} \approx 2 H^{(q-3)/4} R^2 
\end{equation}
which can be integrated to yield approximate (large $R $ ) solutions.

\subsection{1/N Corrections to the Fuzzy Funnel.}
We are interested in the corrections to the funnel solutions we have found, particularly those arising from the
symmetrised trace prescription. In flat space the funnel is a BPS configuration and thus insensitive to any corrections
to all orders. In curved space we have seen that the funnel solution will not generally correspond to a BPS configuration as
the bulk supersymmetries will be broken. Using (\ref{eq:str_corrections}) we can calculate
the leading $1/N$ corrections to the Hamiltonian. As usual it is convenient to introduce the following expressions to 
simplify the results
\begin{equation}
\alpha = e^{-\phi}\sqrt{g_{00}g_{zz}}, \hspace{0.5cm} \beta = \sqrt{1+4\lambda^2 C R^4 g_{xx}^2}, \hspace{0.5cm} 
\gamma = \sqrt{1+\lambda^2 C R^{\prime 2}g_{xx}g_{zz}^{-1}-e^2}, \nonumber
\end{equation}
where we have also introduced the simplification $e^2 = \lambda^2 \varepsilon^2 g_{00}^{-1}g_{zz}^{-1}$.
This allows us to write the first correction to the Hamiltonian, assuming of course that the dilaton term is not
a function of the Casimir
\begin{equation}
\mathcal{H}_1 = \alpha\beta\gamma - \frac{\alpha}{6C}\left\lbrace \frac{2(\beta^2-1)(\gamma^2-1+e^2)}{\beta\gamma}
-\frac{\gamma(\beta^2-1)^2}{\beta^3} - \frac{\beta(\gamma^2-1+e^2)^2}{\gamma^3} \right\rbrace. 
\end{equation}
Now, setting the electric field to zero implies that the correction terms will cancel out to zero when $\beta = \gamma$.  This can actually be seen just by demanding minimisation of $\mathcal{H}_0$, however
we can also see that the correction terms vanish upon implementation of the symmetrised trace.
The minimisation yields a constraint on the curvature which is given by the following
\begin{equation}
R^{\prime 2} = 4R^4g_{xx}g_{zz}.
\end{equation}
In flat space this is just the BPS condition which leads to the simple funnel solution. In certain backgrounds where the
$g_{xx}$ components equal the inverse of the $g_{zz}$ components - for example $Dq$-brane backgrounds - we also recover the
simple funnel solution. However we know that this is only a solution to the equation of motion in the $D3$-brane
background, and so we seem to have found solutions satisfying the minimal energy condition but which do not
solve the equations of motion. In the $NS$5 and $F$-string backgrounds we see that this energy condition coincides
with a solution to the equations of motion, and so we expect those particular funnel solutions to be minimal energy
solutions. This tells us is that the symmetrised trace corrections are zero for configurations which are in their minimal energy 
states. In flat space the minimal energy state coincides with the BPS condition which is why we do not have corrections. In general
the lowest energy configuration may not be BPS but will still receive no corrections from the symmetrised trace.
The general solution consistent with energy minimisation can be written as
\begin{equation}\label{eq:general_soln}
R(\sigma) = \frac{\mp 1}{2\int d\sigma \sqrt{g_{xx}g_{zz}}} \hspace{0.5cm } = \hspace{0.3cm} \frac{\mp 1}{2\int d\sigma f(\sigma)}.
\end{equation}
We expect simple power law behaviour for $f(\sigma) \sim \sigma^n$ and so the solution can be written as
\begin{equation}
R(\sigma) \sim \frac{\mp (n+1) }{2(\sigma^{n+1}-\sigma_0^{n+1})},
\end{equation}
where $n$ can be positive or negative, but not equal to $-1$. The case where $n=1$ corresponds to flat space. In the above expression we have neglected the dimensionality constant coming from the function $f$.
When $n=-1$ the solution reduces to the inverse logarithm solution we find in the $NS$5-brane background.
Note that when $n$ is negative we do not obtain funnel solutions as the radius of the fuzzy sphere never diverges, instead it monotonically increases with the distance from the
sources. This indicates that these solutions do not expand into higher dimensional branes, and will not have an Abelian world-volume description.

Even though the funnel configuration appears to satisfy the energy minimisation condition, the energy itself still has dependence
on the location of the funnel in the throat through the $\alpha$ term. For the three cases where we find explicit brane solutions,
namely the $D3$, $NS$5 and $F$-string backgrounds this term reduces to unity. In the $D5$-brane background we see that 
$\alpha \propto 1/\sigma$ and so the solution minimises its energy when it is far from the sources and thus well 
approximated by the simple funnel solution. The $D1$-brane background yields $\alpha \propto \sigma^3$ and so the
funnel is only a solution when it is on top of the background branes, which is where our effective action will no longer
be valid. This perhaps explains why we were unable to find analytic solutions to the funnel equation of motion.
We should note at this point that $g_s \to 0$ with $\sigma$ in the $D5$ case, implying that the tension of the branes will
become infinite and again our action will be invalidated. In the $D1$ case we see that the coupling becomes strong
as $\sigma \to 0$, therefore the tension of the branes is small but our assumption that $g_s N <1$ must be violated.
It appears that both these backgrounds cause the effective action to break down and so we cannot trust our solutions
except at large $\sigma$, where the background is essentially flat and we recover our simple funnel solution.
The reason why this is not the case in the $NS$5-brane background is because their tension goes as $1/g_s^2$, and so the
coincident brane solution has a much larger mass than the $N$ $D$-strings.

Setting aside the minimal energy condition for a moment we can make some observations about the energy of the funnel including the
leading order correction terms. Firstly we consider the case $R^{\prime}=0$ corresponding to no curvature. The energy can be written as
\begin{equation}
\mathcal{H}_1 = \alpha \beta \sqrt{1-e^2}\left(1+\frac{\alpha(\beta^2-1)^2}{6C\beta^4} \right).
\end{equation}
Clearly when $\beta^2 = 1$ there will be no corrections to the energy, a condition that can be satisfied either by taking $R \to 0$ or
 $g_{xx}^2 \to 0$. The
first condition corresponds to no curvature, with the strings located at an infinite distance away from the background source.
 The second condition is the more interesting
as it generally implies that $\sigma \to 0$, or that the strings are located at the source. The resultant energy for the strings 
is then determined by $\alpha$ - provided we have a sub-critical electric field, 
and so we see that minimising $\alpha$ is equivalent to minimising the energy.
We can also consider the case where we take $R=0$, to see the effect this has on the energy and its corrections. The resultant expression becomes
\begin{equation}
\mathcal{H}_{1} = \alpha \gamma \left(1+\frac{\lambda^2 R^{\prime 2}g_{xx}g_{zz}^{-1}}{6\gamma^4} \right).
\end{equation}
Again we see that the correction term vanishes if we demand the curvature to be zero, or alternatively we can set $g_{xx}g_{zz}^{-1} \to 0$
 either as a product or 
individually, which basically implies that $\sigma \to 0$ as usual. We see once more that $\alpha$ plays the dominant role in determining the energy, and that if this term
can vanish then so can the energy. This helps to explain why we cannot obtain analytic solutions for the $D5$ and $D1$-backgrounds, as in these cases the $\alpha$ term is
a function of $\sigma$ which implies that the energy will either diverge, or tend to zero with $\sigma$, depending on the dimensionality of the source branes. Therefore the 
energy is dependent upon the space-time variables. For the $D3$, $NS$5 and $F$-string
backgrounds we find that $\alpha = 1$ and thus it is the shape of the funnel itself which dictates the minimal energy configuration.

\subsection{Time Dependence and Dualities.}
In the time dependent case we again use the conservation of the energy-momentum tensor to obtain the equation of motion
\begin{equation}\label{eq:dynamic_eom}
\dot{R}^2 = \frac{g_{00}(1-g_{00}^{-1}g_{zz}^{-1}\lambda^2 \varepsilon^2) A}{\lambda^2 C g_{xx} \bar{g}_{00}\bar{g}_{zz}(1+4\lambda^2 C R_0^4 \bar{g}_{xx}^2)(1-\bar{g}^{-1}_{00}\bar{g}^{-1}_{zz}\lambda^2 \varepsilon^2)},
\end{equation}
where the coefficient $A$ is written as follows
\begin{equation}
A=-(e^{2(\phi_{0}-\phi)}g_{00}g_{zz}(1+4\lambda^2 C R^4 g_{xx}^2)(1-g_{00}^{-1}g_{zz}^{-1}\lambda^2 \varepsilon^2)-\bar{g}_{00}\bar{g}_{zz}(1+4\lambda^2 C R_0^4 \bar{g}_{xx}^2)(1-\bar{g}_{00}^{-1}\bar{g}_{zz}^{-1}\lambda^2 \varepsilon^2)) \nonumber.
\end{equation}
In deriving this expression we have imposed the initial conditions that $R(t=0)=R_0$ when $\dot{R}=0$, and the metric components at this initial point have been
denoted by a bar. Note also the factor of $e^{\phi_0}$ in the solution which reflects the initial value of the dilaton subject to these boundary conditions.
In fact this equation is remarkably similar to the static
one, which can be calculated to yield
\begin{equation}
R^{\prime 2} = \frac{g_{zz}(1-g^{-1}_{00}g^{-1}_{zz}\lambda^2 \varepsilon^2) B}{\lambda^2 C g_{xx}\bar{g}_{00}\bar{g}_{zz}(1+4\lambda^2CR_0^4\bar{g}_{xx}^{2})(1-\bar{g}_{00}^{-1}\bar{g}_{zz}^{-1}\lambda^2 \varepsilon^2)},
\end{equation}
where the coefficient $B$ turns out to be simply  $-A$ 
If we consider the case where the $D$-string is located far from the sources in flat Minkowski space, the metric
components and the dilaton can be set to unity.
in this limit the two equations of motion reduce to
\begin{eqnarray}
R^{\prime 2} &=& \frac{4(1-\lambda^2 \varepsilon^2)(R_0^4-R^4)}{1+4\lambda^2CR_0^4} \\
\dot{R}^2 &=& \frac{4(1-\lambda^2 \varepsilon^2)(R^4-R_0^4)}{1+4\lambda^2C R_0^4}, \nonumber
\end{eqnarray}
which are clearly invariant under the following invertible world-sheet transformation $t \to i \sigma$, which is nothing more than Wick rotation.
If we re-write these equations using dimensionless variables as in (\ref{eq:dimensionless}), introducing a similar transformation on the $\sigma$ coordinate,
then we find that the two equations of motion are related via $\dot{z}=iz^{\prime}$. Therefore knowledge of one of the solutions (\ref{eq:flat_solution}) automatically
implies knowledge of the other solution as follows
\begin{eqnarray}
z(\tau) &=& \pm z_0 {\rm JacobiCN} \left\lbrack \frac{\sqrt{2(1-e^2)}\tau z_0}{\sqrt{1+z_0^4}}, \frac{1}{\sqrt{2}} \right\rbrack \\
z(\sigma) &=& \pm \frac{z_0}{ {\rm JacobiCN} \left\lbrack \frac{\sqrt{2(1-e^2)}\sigma z_0}{\sqrt{1+z_0^4}}, \frac{1}{\sqrt{2}} \right\rbrack}.\nonumber
\end{eqnarray}
In the last line we have used one of the various properties of elliptic functions. As discussed in an earlier paper \cite{costis},
the last equation defines a periodic array of $D3$/$\bar{D3}$-branes connected by the fuzzy $D1$-funnels.
There are two important comments to be made at this point. Firstly that the equation of motion for a collapsing fuzzy
sphere is the same as that of a time-dependent funnel in Minkowski space. Secondly the world-sheet transformation
we employed on the equation of motion has a geometric interpretation. Instead of performing a Wick rotation on the time
variable, we can instead identify $\tau$ with $\sigma$ provided we also send $z \to 1/z$. Using the definition of
the elliptic function we can easily verify that this is true. Therefore we have a concrete example of the so called
large/ small duality \cite{costis} that pervades all string theories, as a collapsing fuzzy sphere of radius $R$ is dual to a
brane-anti-brane array with interpolating funnel solutions of maximal radius $1/R$. 

In the more general case it is clear to see that we recover the static equation from the time dependent one by 
performing the following transformation
\begin{equation}
\varepsilon \to 0, \hspace{1cm} t \to i\sigma, \hspace{1cm} g_{00} \to g_{zz}.
\end{equation}
This corresponds to a Wick rotation on the worldsheet and a space-time transformation in the bulk, and is therefore
a highly non-trivial symmetry. However we can see that the transformation is not invertible, unlike in Minkowski space, 
due to the $\sigma$ dependence of the metric components. If we start with the static equation and rotate the spatial
coordinate such that $\sigma \to -i\tau$, then the metric components (as well as the curvature term) become time 
dependent - corresponding to some form of time dependent background \footnote{Unfortunately these are not the spacelike D-brane supergravity solutions constructed in \cite{peet}.}.
 If we take this solution and then Wick rotate the time variable again we recover the spatial dependent equation.
Thus it appears there is a mapping from the time dependent equation to the static one, but not vice-versa. The static equation
is invariant under a double Wick rotation, which appears to be the only automorphism of that particular equation.
This implies that the large/small duality is broken in this instance by the presence of curved spacetime, which we ought to expect
since the time-like and space-like Killing vectors cannot be rotated into one another due to the additional spatial dependence of the metric components. In flat space
the metric, and therefore the Killing vectors, are invariant under Wick rotation and so the field theory solutions ought to respect this symmetry. 

There is, however, a particularly interesting transformation in curved space when the metric
components $g_{00}$ and $g_{zz}$ are inverses of each other - as in the case for $Dq$-brane backgrounds in the near 
horizon limit.
Writing the harmonic function in terms of the dimensionless distance variable $\tilde z$
\begin{equation}
H \sim \frac{1}{\tilde z^{7-q}},
\end{equation}
then it is straightforward to see that the transformation to the static equation
is nothing more than T-duality, taking $\tilde z \to 1/\tilde z$.

One further comment should be made here with regard to the interpretation of the dynamical solution. In the Minkowski
limit we saw that the time dependent funnel solution yielded the same equations of motion as the collapsing fuzzy sphere.
This led \cite{costis} to postulate the existence of a duality between contracting fuzzy spheres and funnels. In curved space we see that this
interpretation is no longer valid, since the equations of motion coming from the collapsing fuzzy sphere are different - as
shown in section 2.
\section{The Dual Picture - $D3$ world-volume theory.}
Our work on constructing funnel solutions in curved space has yielded some interesting results. At this stage we would like to check our assumption
that the funnels do in fact lead to the emergence of $D3$-branes, which can be done in the dual $D3$ world-volume theory. We begin with
the effective action for a solitary $D3$-brane in a general background with vanishing Kalb-Ramond two form
\begin{equation}
S = -\tau_3 \int d^4 \zeta e^{-\phi}\sqrt{-det(G_{ab}+\lambda F_{ab})},
\end{equation}
where $G_{ab}$ is the pullback of the background metric to the world-volume and $F_{ab}$ is the $U(1)$ field strength as usual.
The $D$-strings in this theory will appear as magnetic monopoles on the $D3$-brane, thus we must ensure a non-trivial magnetic field is turned on.
We choose this to be $F_{ab}=\epsilon_{abc}B_c$, with roman indices running over the world-volume. Finally we must also ensure that one of the transverse
scalars - $\sigma $ is excited. As usual we neglect higher derivative terms in the DBI action, and 
employ the use of static gauge. The result for the static solution is as follows
\begin{equation}
S = -\tau_3 \int d^4{\zeta} e^{-\phi} \sqrt{g_{00}g_{xx}^{3}(1+\lambda^2 g_{zz}g_{xx}^{-1}(\vec{\nabla}\sigma)^2+\lambda^2g_{xx}^{-2}\vec{B}^2+\lambda^4g_{zz}g_{xx}^{-3}(\vec{B}. \vec{\nabla}\sigma)^2)}.
\end{equation}
It should be noted that the scalar field has canonical dimension of $L^{-1}$, which we need to be careful of when
interpreting our solutions - particularly when trying to show that this is indeed the dual picture configuration.
The equation of motion for the transverse scalar is complicated in curved space, and not readily amenable to analytic
solutions. Thus we will attempt to find the spike profiles by searching for configurations which minimise the energy,
a tactic which worked for several backgrounds in the non-Abelian case, where the energy minimisation condition corresponded
to the equations of motion.
The energy density in the static simply equals $-\mathcal{L}$ therefore we may write
\begin{eqnarray}
\mathcal{H} &=& \tau_3 \int d^3{\zeta} e^{-\phi} \sqrt{g_{00}g_{xx}^{3}(1+\lambda^2 g_{zz}g_{xx}^{-1}(\vec{\nabla}\sigma)^2+\lambda^2g_{xx}^{-2}\vec{B}^2+\lambda^4g_{zz}g_{xx}^{-3}(\vec{B}. \vec{\nabla}\sigma)^2)} \nonumber \\
&=& \tau_3 \int d^3\zeta e^{-\phi} \sqrt{g_{00}g_{xx}^3}\sqrt{\lambda^2 |\sqrt{g_{zz}g_{xx}^{-1}} \vec\nabla\sigma \pm g_{xx}^{-1} \vec{B} |^2 + (1 \mp \lambda^2 g_{zz}^{1/2}g_{xx}^{-3/2} \vec B. \vec\nabla \sigma)^2}, \nonumber
\end{eqnarray}
where in the last line we have written the determinant as the sum of two squares. 
We see that there is an energy bound given by
\begin{equation}\label{eq:energy_bound}
\mathcal{H} \ge \tau_3 \int d^3 \zeta e^{-\phi}\sqrt{g_{00}g_{xx}^3} |1 \mp \lambda^2 g_{zz}^{1/2}g_{xx}^{-3/2} \vec B. \vec\nabla \sigma |,
\end{equation}
which is saturated provided that the $\sigma $-field satisfies the following constraint
\begin{equation}\label{eq:bfield}
\vec B = \mp \vec \nabla \sigma \sqrt{g_{xx}g_{zz}},
\end{equation}
which can be seen to reduce to the usual flat space constraint $\vec B = \mp \vec \nabla \sigma$ as required.
The expression for the energy bound (\ref{eq:energy_bound}) seems to be the sum of two terms where the second one is 
topological in nature. We wish to show that this expression has a simple interpretation in terms of the energy of the
$D3$-brane and the energy of a warped spike solution. We will write the first term as follows
\begin{equation}
\mathcal{H}_{D3} = \tau_3 \int d^3 \zeta e^{-\phi} \sqrt{g_{00}g_{xx}^3}.
\end{equation}
Now in flat space the energy of the $D3$-brane is simply $\tau_3 \int d^3 \zeta$, however as we are in a generic curved
background we must also include the contribution from a non-trivial dilaton. This means the energy
is modified to become $\tau_3 \int d^3 \zeta e^{-\phi} $ which is exactly the equation we wrote down for the
energy of a warped $D3$-brane. Thus our intuition about the first term is correct, namely that it corresponds to the energy
of the brane in curved space. The second term is a simple extension of the BIon spike solution, generalised to a curved background.

We return now to (\ref{eq:bfield}) which gives us 
important information about the profile of the spike solution.
It is clear that the second term here is a total derivative if the $B$ field satisfies 
the modified Gauss law equation
\begin{equation}\label{eq:gauss}
\vec \nabla . (\sqrt{g_{00}g_{zz}} e^{-\phi}\vec B) = 0.
\end{equation}
This modification of the Gauss law appears to be due to red-shifting of the magnetic field for an observer in the UV end
of the background geometry. Such red-shifting effects are common in warped metrics.
Under these circumstances the second term would then be determined by the boundary values of $\sigma(r) $ 
and so we would find a contribution to the energy proportional to  $\sigma(r=\infty) - \sigma(r=0)$ 
which could be interpreted as the energy of a string stretching along the $\sigma $ direction .i.e. the
$D$-strings of the non-Abelian theory.

In the case of background $NS5$ branes or $F$-strings, which are both charged under the $NS$ field,
it is easy to check that (\ref{eq:gauss})
reduces to the familiar flat-space gauss constraint due to the cancellation with the metric components
$\vec \nabla .\vec B = 0$. In the case of $Dq$-brane backgrounds, which are charged under the $RR$ fields, such a cancellation between the dilaton and 
metric components does not occur and the Gauss law condition reduces to
\begin{equation}\label{eq:gauss2}
\vec \nabla . (e^{-\phi} \vec B) = 0.
\end{equation}
We wish to solve the general spike solution using (\ref{eq:bfield}). In general we may expect a power series solution for
the metric functions which will be given by $f(\tilde\sigma)$, where $\tilde\sigma$ refers to the physical coordinate distance.
Note that $\sigma$ is related to the physical distance via $\tilde\sigma = l_s^2 \sigma$. As in the non-Abelian section we will take
$f(\tilde\sigma) \sim \tilde \sigma^n$, where $n$ can be positive or negative but not unity.
It will be convenient to switch to spherical coordinates in which case the magnetic field will only have a radial dependence, and we will
take the traditional ansatz for the field to be
\begin{equation}
B = \frac{\pm Q}{4\pi r^2},
\end{equation}
where $Q$ corresponds to the magnetic charge of the field. Equating both sides of (\ref{eq:bfield}) gives us the physical solution for the spike
\begin{equation}
\tilde \sigma^{n+1} - \tilde \sigma_0^{n+1} \propto \frac{\pm Ql_s^2(n+1)}{4\pi r},
\end{equation}
where we have neglected a dimensionality factor which makes $f(\tilde \sigma)$ dimensionless. With reference to the general solution
on the non-Abelian side (\ref{eq:general_soln}) in physical coordinates we find
\begin{equation}
\sigma^{n+1}-\sigma_0^{n+1} \sim \frac{\pm \pi N l_s^2 (n+1)}{r}.
\end{equation}
If we demand that both of these solutions are equal - to leading order in $N$ - we need to impose the following quantisation condition on the magnetic charge, 
namely $Q = 4\pi^2 N$. This condition, with the appropriate choice of sign, ensures that the equations from the non-Abelian and Abelian theories are the same
in an arbitrary background. The $n=-1$ case, which arises in the fivebrane backgrounds, will give rise to a logarithmic funnel profile and not the simple
power law solution. 

In the specific case of the $NS$5-brane background we find that the spike solution from the Abelian action, $\tilde{\sigma}(r)$ satisfies the 
following equation
\begin{equation}\label{eq:D3side}
-\frac{1}{r} - \frac{4\pi \sqrt{M}}{Q}  \sqrt{\frac{\tilde{\sigma}^2}{l_s^2 M} +1} +\frac{2\pi \sqrt{M}}{Q} \ln{\left(\frac{\sqrt{\frac{\tilde{\sigma}^2}{l_s^2 M} 
+1}}{\sqrt{\frac{\tilde{\sigma}^2}{l_s^2 M} -1}} \right)} = c 
\end{equation}
where $c$ is a arbitrary constant of integration. 
In the throat approximation where $\frac{\tilde{\sigma}^2}{l_s^2 M} \ll 1 $ this equation can be solved explicitly
for the spike profile 
\begin{equation}\label{eq:spike}
\tilde \sigma = \tilde\sigma_0 \exp \left(\frac{- Ql_s}{4\pi r \sqrt{M}}\right) = \tilde\sigma_0 \exp \left(\frac{-\pi l_s N}{\sqrt{M}r} \right),
\end{equation}
More generally the complete solution above in (\ref{eq:D3side}) can be seen to be exactly equivalent to the solution for the fuzzy funnel discovered 
on the non-Abelian side in (\ref{eq:nonabside}) with an appropriate definition of the constant $c$ in terms of the D3-brane location parameter
$\sigma_0 $ and using the quantisation of magnetic charge $Q$ found earlier.

Now in flat space the fact that a spike profile saturates the energy bound is normally sufficient to argue that such a profile solves 
the equations of motion. However in the case where there is a throat present due to the NS5 source branes, this is not the case. 
From equation (\ref{eq:bfield}) with $g_{xx} $ and $g_{zz}$ appropriate to the throat geometry, we can scale  
$\tilde \sigma \to l\tilde \sigma$  and still satisfy this equation.  However under the same scaling, the energy of the warped 
$D3$-brane scales like
\begin{equation}
\mathcal{H}_{D3} \to l \mathcal{H}_{D3}
\end{equation}
and so the energy of the brane can now be reduced by sending $l \to 0$, indicating that the $D3$-brane - or funnel solution on 
the non-Abelian side - will be unstable. This shows that the static spike profile (\ref{eq:D3side}) is unstable and wants to decay. 
Thus by considering a time-dependent profile rather than static, we can find a solution to the equations of motion.  

In general looking for analytic $t$ and $r$-dependent solutions to the equations of motion looks very difficult.
However assuming the throat approximation, a simple solution, which describes the motion of the funnel as a whole, can be obtained by 
using separation of variables.
Such a solution can be expressed as  $F(t)\tilde \sigma(r)$, where 
we have introduced
a dimensionless time-dependent profile, $F(t)$ for the spike. It is easy to see that $F(t)$ drops out of 
 eq(\ref{eq:bfield}) so that $\tilde{\sigma}(r) $ still describes a static spike profile as in eq(\ref{eq:spike}). 
$F(t) $ is determined by demanding $F(t)\tilde{\sigma}(r) $ solves the complete equations of motion. 
 We find
that the energy density of the brane reduces to the simple form
\begin{equation}
{\cal E} = \tau_3 V_3 \frac{F\tilde \sigma}{\sqrt{Ml_s^2}} \left| 1 \pm \frac{\lambda^2}{l_s} \sqrt{M} \vec{B}.\vec{\nabla} \ln \left(\frac{\tilde \sigma}{\tilde \sigma_0}\right)\right| \left(1-\frac{\lambda^2 M \dot{F}^2}{l_s^2F^2} \right)^{-1/2},
\end{equation}
where $V_3$ is the volume element of the $D3$-brane. Demanding  the conservation of energy 
(equivalent to solving the equations of motion) we can solve for 
for $F(t)$, noting that the absolute value of the second term is independent of time. The solution can be seen to yield
\begin{equation}
\frac{1}{F(t)}=\frac{1}{F_0}\cosh \left(\frac{tl_s}{\lambda \sqrt{M}} \right),
\end{equation}
where $F_0$ is the initial condition on the profile. There are two important comments to make here. Firstly that the solution appears
to be valid for any point on the world-volume, even at the location of the monopole $r=0$. Secondly the solution for the profile
is exactly the same functional form as that of a $D3$-brane with no magnetic flux in the same background, as shown
by Kutasov in ref.\cite{time_dependence}. This suggests that the BIon spike will not feel any tidal forces due to the gravitational attraction of the fivebranes.
We may now write the full solution to the equation of motion (again in the throat approximation) as follows
\begin{equation}
r(\sigma, t) = \frac{N\pi \lambda l_s}{\lambda \sqrt{M} \ln \left(\frac{\tilde \sigma}{\tilde \sigma_0} [1+e^{-tl_s/\sqrt{M}\lambda}]\right)+tl_s + \lambda\sqrt{M}\ln(2)},
\end{equation}
which we can simplify by considering the solution at late times - and neglecting the constants arising from the initial conditions
\begin{equation}\label{eq:eom_approx}
r(\sigma, t) \sim \frac{N \pi \lambda l_s}{\lambda \sqrt{M}\ln\left(\frac{\sigma}{\sigma_0}\right)+tl_s}
\end{equation}
which shows that the radion field is proportional to $1/t$ in this limit. We now want to consider how this appears on the
non-Abelian side, however we note that even when we include time dependence in the action the equations of motion are
highly non trivial and do not yield a simple analytic solution. We should check that the solution (\ref{eq:eom_approx})
is actually a solution of the theory.
We again factorise the scalar field into a time dependent piece and a spatial piece and make the ansatz
\begin{equation}
R(\sigma, t) = \frac{1}{2\sqrt{Ml_s^2}\ln\left(\frac{\sigma}{\sigma_0}\right)+BF(t)},
\end{equation}
where $B$ is some arbitrary constant. It can easily be seen that $R' = 2R^2 \sqrt{H}$ and $\dot{R}=B\dot{F}R^2$, where $H$ is the
usual harmonic function for the $NS$5-brane solution. If we substitute these two equations into the energy density equation
for the fuzzy funnel we obtain
\begin{equation}
\mathcal{H} = \frac{\tau_1 V_1 N (1+4\lambda^2 C R^4)^{3/2}}{\sqrt{1+4\lambda^2 C R^4-\lambda^2 C B^2 R^4 \dot{F}^2}},
\end{equation}
which must be conserved in time. This requires that the $\dot{F}$ term must vanish from the expression. The simplest
solution is to take $\dot{F}=0$, however this implies that $F$ is constant in time and so we are just introducing
a constant shift into the equation of motion. A non-trivial solution can be obtained by setting $\dot{F}^2 B^2=4$, which
has the solution $F(t)=2t/B$. This reproduces the same functional form for the equation of motion as we derived from
the Abelian theory, however we need to check the interpretation of the resultant expression for the energy density, which can
be seen to yield
\begin{equation}
E \to \tau_1 V_1 N (1+4\lambda^2 C R^4)^{3/2}.
\end{equation}
Expanding the solution we can see the first term corresponds to the energy density of $N$ coincident $D$-strings, as 
we would expect. The higher order terms correspond to non-linearities arising from the fuzzy funnel solution representing
the warping of the $D$-strings in the transverse space. Thus we argue that this ansatz for the equation of motion is 
a solution of the theory as we are left with the minimal energy configuration.
Therefore both solutions agree at late times. Furthermore it was argued in \cite{time_dependence} that
we can trust the macroscopic description even deep in the IR end of the geometry provided that the energy of the brane
is large enough. Therefore we expect our solution to capture the vast majority of the evolution of the system.
Of course, our analysis is based upon the fact that we are ignoring the back reaction upon the geometry. Again this
requires fine tuning of the various parameters in the theory to accomplish this. Hopefully using the prescription for
the symmetrised trace at finite N will alleviate this problem entirely.

Similar analyses can be carried out for both the $F$-string and $Dq$-brane backgrounds. The static spike profile in the 
$F$-string background, obtained by solving (\ref{eq:bfield})  is consistent with the static funnel profile obtained in the same 
background on the non-Abelian side. The same scaling argument about such static solutions being unstable, as discussed in the $NS5$
case above, is not naively applicable here. What we can verify is that at least in the static case, the equation for the spike on the
Abelian $D3$-world volume side and the fuzzy funnel on the non-Abelian side agree. 

Finally we discuss the situation for $Dq$-background geometry with $q \neq 3$. Here things are obviously more complicated
due to the red-shift of the magnetic field. However we can use some intuition from our knowledge of the Abelian theory
to understand the physics. It is known that for supersymmetry to be preserved we require the $D3$-brane to be embedded in
either a $D3$-brane or $D7$-brane background \cite{branonium}. In this case the funnel solution will be completely solvable. For all other
brane backgrounds the supersymmetry is broken, and the $D3$ feels a gravitational potential drawing it toward the background branes.
Thus our static funnel solution will not be compatible with the full equations of motion, and so we would require a time dependent
ansatz. Interestingly in the $D5$-brane background we know that open string modes stretching between the funnel and the 
source branes will become tachyonic at late times, potentially distorting the funnel.

\section{Higher Dimensional Fuzzy Funnels.}
We can generalise the non-Abelian results we have obtained to the higher dimensional theory using the work in \cite{costis, thomas} as our basis.
This means we are considering the fuzzy $\mathcal{S}^{2k}$ spheres, which are labelled by the group structure of $SO(2k+1)$ in ten dimensions. This will obviously imply
that we require $2k+1$ transverse scalars in the DBI action, where $k \le 3$ and the funnels are now blowing up into $ n D(2k+1)$-branes in an arbitrary background.
Of course the higher number of transverse directions will impose serious constraints upon the dimensionality of the possible background sources, in many cases we will
be left with unphysical situations such as type IIA, or potentially non braney solutions. The geometry of these higher dimensional fuzzy spheres is interesting to study in its own right, for example we know that
the fuzzy $S^6$ can be written as a bundle over the classical six-sphere \cite{ho}. In the classical limit we find that the fibre over the sphere belongs to the group $SO(6)/U(3)$, which
implies that constructing a dual picture is non-trivial. The geometrical analysis is revealing as we can calculate the charge of the branes directly from the base space.
The general topology of our higher dimensional funnel configuration will now be $\mathbb{R}\times S^{2k}$, and we must modify our gauge group ansatz to read
\begin{equation}
\phi^i = \pm RG^i,
\end{equation}
where the $G^i$ matrices satisfy $G^iG^i = C_k \mathbf1_N$ and lie in the irreducible representation of the particular gauge group. The Casimir in this case will be labelled by a $k$ index
so that we know which group structure it conforms to. These generators will arise from the action of gamma matrices on
traceless, symmetric $n$-fold tensor products of spinors \cite{costis} and generally do not form a closed Lie algebra.
 The relationship between $N$ and $n$ means that the dual picture is far more complicated.
For example in the $k=2$ case we know that the $D$-strings blow up to form several $D5$-branes, which have a non-trivial second Chern Class on the world-volume. This
makes the dual picture difficult to analyse and we will not do it in this note - but see \cite{constable, cook} for a more detailed derivation of the $D1-D5$ and $D1-D7$
solutions in flat space. The general relationship between the physical distance and the scalar field ansatz can be written as follows
\begin{equation}
r = k \sqrt{C_k} \lambda R,
\end{equation}
which is similar to the $SU(2)$ case, except there is no ambiguity over the choice of sign, and we emphasise that the Casimir will be dependent upon the number
of higher dimensional branes in the funnel solution.

The generalisation of the non-Abelian action to leading order is expected to be given by
\begin{equation}
S=-\tau_1 \int d^2 \sigma N e^{-\phi} \sqrt{g_{00}g_{zz}(1+\lambda^2 C_k g_{xx}g_{zz}^{-1}R^{\prime 2})}(1+4\lambda^2 C_k R^4 g_{xx}^2)^{k/2},
\end{equation}
and therefore with our usual rescaling of the tension we can find the spatial component of the energy momentum tensor
\begin{equation}
T_{\sigma \sigma} = \frac{e^{-\phi}\sqrt{g_{00}g_{zz}}(1+4\lambda^2C_kR^4g_{xx}^2)^{k/2}}{\sqrt{1+\lambda^2 C_k R^{\prime 2} g_{xx}g_{zz}^{-1}}}.
\end{equation}
Our work in the lower dimensional case has shown that we can obtain solutions to the equations of motion, consistent with the energy minimisation principle, when the $\alpha$ term
is constant. If we assume that this is true for our background metric then we can write the general equation of motion for the funnel as follows
\begin{equation}
R_k^{\prime 2} = \frac{g_{zz}}{\lambda^2 C_k g_{xx}}\left( (1+4\lambda^2 C_k R^4 g_{xx}^2)^k -1 \right).
\end{equation}
A quick check shows that with $k=1$ the solution reduces to $R_1^{\prime 2} = 4 R^4 g_{xx}g_{zz}$ as expected from our efforts in the preceding sections. Of course setting $\alpha$ to be constant
also imposes additional constraints on the possible supergravity backgrounds that exist.
Interestingly the higher dimensional solutions will all have a variant of this solution as their lowest order expansion in $\lambda$.
The $k=2$ and $k=3$ solutions can be written as follows
\begin{eqnarray}
R_2^{\prime 2}&=&8 \left(R^4g_{xx}g_{zz}+2\lambda^2 C_2 R^8 g_{xx}^3g_{zz} \right) \\
R_3^{\prime 2} &=& 12\left( R^4g_{xx}g_{zz}+4\lambda^2 C_3 R^8 g_{xx}^3g_{zz}+\frac{16}{3}\lambda^4 C_3^2R^{12}g_{xx}^5g_{zz} \right) \nonumber
\end{eqnarray}
which shows that there are apparent recursive properties for these equations.
Note that these expression agree exactly with the ones derived in \cite{constable, cook} when taking the flat space limit, where these results were obtained
via minimisation of the energy and found to be perturbatively stable. Clearly we do not expect this to be the case in a general background due to the additional
$\sigma$ dependence of the metric components.

In general these equations are difficult to solve, but can in principle be written in terms 
of elliptic functions. We will try and make some progress by assuming trivial solutions for the $g_{xx}$ components which can be absorbed into a redefinition of $R$, and power law behaviour for the $g_{zz}$ components.
In the $k=2$ case we can find approximate solutions to the equation of motion. In the large $R$ region, the second term is dominant and a quick integration yields the following solution
\begin{eqnarray}
R_2(\sigma) &=& \left(  \frac{\mp 1}{4\lambda \sqrt{C_2}(\sigma^{m+1}-\sigma_0^{m+1})} \right)^{1/3} \hspace{2cm} m \ne -1 \\
R_2(\sigma)&=& \left( \frac{\mp 1}{4\lambda\sqrt{C_2}\ln(\sigma/\sigma_0)}\right)^{1/3} \hspace{3.1cm} m=-1. \nonumber
\end{eqnarray}
Note that $m=0$ corresponds to the flat space limit and agrees with the solution in \cite{constable}. When $R$ is small, the solution is dominated by
the leading term and we recover the usual funnel solution derived in previous sections. Clearly this implies the existence of an interpolating region where the solutions
cross over from one another. Upon equating the two terms we find that the cross over occurs at
\begin{equation}
R_{cr} \sim \left(\frac{1}{2\lambda^2 C_2} \right)^{1/4},
\end{equation}
which implies, in physical coordinates, that $r >> l_s$. Moving on to the $k=3$ case, we find it complicated by the appearance of an extra term. Of course, in the large $R$ limit
this will be the dominant contribution to the integral and we find a similar solution to the one sketched out above with the power now being $1/5$ rather than $1/3$ , and the
dependence on $\lambda$ and $C$ will also be slightly altered. The crossover in this case will happen at the point
\begin{equation}
R_{cr} \sim \left(\frac{3}{8\lambda^2 C_3} \left\lbrace 1+ \sqrt{\frac{7}{3}} \right\rbrace \right)^{1/4},
\end{equation}
which will again imply that the physical distance is much larger than the string scale.
The general conclusion here is that higher dimensional fuzzy spheres lead to funnel solutions which are modified version of the lower dimensional ones, although we ought to
bear in mind that these solutions are potentially only valid in flat space as physical brane sources satisfying the background constraints may not exist. The general behaviour for the funnel in the
large $R$ limit can be seen to be
\begin{equation}
R \sim \sigma^{-(m+1)/(2k-1)},
\end{equation}
and so the higher dimensional effects play a more important role as $\sigma \to \sigma_0$.

We now switch our attention to the leading order $1/N$ corrections for the general fuzzy funnel. As usual we choose to work in terms of the variables $\alpha, \beta, \gamma$, where
now $\beta$ is the general function for arbitrary $k$. The leading order correction can be calculated to give
\begin{equation}
\tilde{\mathcal{H}_1} = \alpha \beta \gamma \left\lbrace1-\frac{1}{3\gamma C_k} \left(\frac{k(\gamma^2-1)(\beta^{2/k}-1)}{\gamma \beta^{2/k}}-\frac{(\gamma^2-1)^2}{2\gamma^3}+\frac{k\gamma(k-2)(\beta^{2/k}-1)^2}{2\beta^{4/k}} \right) \right\rbrace \nonumber
\end{equation}
which clearly reduces to the standard expression when $k=1$. This is actually valid for $k=4$ provided we take the flat space limit.
Now we see that in general the correction terms  will be non-zero, even if we 
assume the funnel configuration where $\beta=\gamma$. This is actually reminiscent of the flat space solutions where the higher dimensional
fuzzy funnels are corrected under the symmetrised trace. Taking $k=2$ for example we find that the corrected energy becomes
\begin{equation}
\tilde{\mathcal{H}_1} = \alpha \beta \gamma \left\lbrace 1-\frac{1}{3\gamma C_2}\left(\frac{2(\gamma^2-1)(\beta-1)}{\gamma \beta}-\frac{(\gamma^2-1)^2}{2\gamma^3} \right) \right\rbrace,
\end{equation}
which implies that the correction terms only vanish for $\gamma^2=1$. The non-trivial solution to this implies that $R^{\prime}=0$, or that the radius of the sphere is a constant function of $\sigma$.
Furthermore we see that the correction term will always be positive, therefore the higher order corrections reduce the energy and so we expect the solution to be unstable. It is only the
$D1-D3$ funnel which is the lowest energy configuration in an arbitrary background.
\section{Discussion.}
We have further investigated the large $N$ limit of the non-Abelian DBI in curved backgrounds. As anticipated, the presence
of electric flux on the coincident $Dp$-branes will never prevent the fuzzy sphere from collapsing toward zero
size, however we may expect quantum effects to become more important as the branes near one another. In the event that
the field saturates its maximum bound the fuzzy sphere is static, however we should interpret this as a failure of the
non-Abelian action rather than a physical condition.

More importantly, we have investigated the fuzzy funnel solution in the same background and found a variety of differing
behaviour depending on the exact form of the background metric. For those cases where funnels can exist we constructed
the most general class of solutions, which were either power law or logarithmic in profile. The non-trivial facts
about this construction can be summarised as follows. By demanding that the strings expand into a $D3$-brane in the
throat geometry, we have constructed a geometry where a finite string ends on one side of the brane whilst an infinite
string starts from the other side. However these solutions are apparently related by a $\sigma \to 1/\sigma$ duality which
affects the energy of the solution.
Another interesting property of these solutions is that the symmetrised trace does not provide corrections to the
geometry of the funnel solution, implying that it is the lowest energy configuration. However, we saw that the solutions to the equations of motion
must be modified to include time-dependence in order to obtain a complete description of the fuzzy funnel. 

We constructed the dual Abelian theory in the same background and found that the equations of motion for the 
BIon spike are indeed dual to the fuzzy funnel profile in the large $N$ limit, provided we impose a certain
quantisation condition upon the monopole charge. This implies that the leading order contributions to the action agree even
in a curved background without requiring modification. Furthermore we see that the same backgrounds which posed us 
problems on the non-Abelian side are also non-trivial from the Abelian side. Again we see that in these situations
the funnel/ BIon solutions reduce to the flat space ones. This implies that the Abelian and non-Abelian actions agree
in the same limit, namely large $\sigma$. This behaviour is unexpected as the two descriptions are usually valid in
different regimes.

In addition we looked at the leading order action where there are time dependent scalar fields, and searched for an 
extension of the large/small dualities between collapsing fuzzy spheres and fuzzy funnels \cite{costis}. Although the latter 
configuration can be reached from the former by Wick rotation, we see that the funnel solution cannot be mapped to the
dynamical one. Instead the funnel solution is invariant under a double Wick rotation, which appears to be the only
automorphism of the equation. Thus the curved background has broken the symmetries present in flat space.

We extended the fuzzy funnel solutions to higher dimensional fuzzy spheres. Although this is possible in principle, the
physics will depend on the existence of specific stable backgrounds. The funnel solutions are extensions of the lower
dimensional one, which have different behaviour as the radius of the fuzzy sphere diverges. However the leading
order $1/N$ corrections coming from the symmetrised trace show that these higher dimensional funnels will be unstable even
in the flat space limit. 

Having constructed these fuzzy funnels it seems only reasonable to consider their physical properties, such as 
electromagnetic scattering, and potential uplifts to M-theory. We hope to return to these issues in a subsequent paper.
There are several puzzling issues to resolve, which we hope will be the subject of future investigation. The first is
related to the $D5$-background solution, which appears to force the funnel to open up in flat space. This solution
should be related to the $NS$5-brane solution via S-duality, however it is clear that the physics of the two solutions
are very different. In fact the latter case appears to yield a nice stable funnel solution which even yields
a time dependent profile. A second issue relates to the breaking of the duality symmetry possessed by the flat space
solutions. We suspect that the background acts to break the general solution symmetry into a connected and 
disconnected part. It would be interesting to understand the underlying geometry of this symmetry in terms of Riemmanian geometry,
and how it is related to the symmetrised trace.

A related issue is the finite $N$ expansion of the action. This has been discussed in \cite{rst}, and certainly 
demands further consideration in the context of our analysis. Moreover recent work \cite{finiten} has conjectured a complete
expansion of the symmetrised trace, opening up the possibility of studying all the finite-$N$ effects. This has important
consequences for the microscopical description of the theory, as in this limit we can also neglect back reaction upon the geometry
by considering smaller values of $N$. The work in this note implicitly assumed that the back reaction could be neglected by tuning
the numbers of branes appropriately. We should certainly be careful about this kind of assumption, and certainly only
consider our solutions to be leading order approximations. In fact this could be very useful for describing brane polarisation
in warped backgrounds, such as that proposed by KKLT, as we have the possibility of realising inflation along the lines of 
\cite{inflaton}, or considering cosmic string networks along the lines of \cite{cosmicstrings}. We leave such work for future study. Once potential drawback to this is that we may have to modify the
finite $N$ action in these backgrounds along the lines of \cite{modifiedDBI}. Again this is something that needs clarification.

Following on from the recent work in \cite{koch} it would be useful to consider perturbative fluctuations of these funnel
solutions especially in light of the apparent relation between the zero modes of the fluctuations and the moduli space
of the fuzzy sphere.

\begin{center}
\textbf{Acknowledgements.}
\end{center}
We wish to thank Sanjaye Ramgoolam and Costis Papageorgakis for illuminating discussions. JW is supported by a
QMUL studentship. This work is in part supported by the EC Marie Curie research Training Network MRTN-CT-2004-512194.

\end{document}